\documentclass{article}

\usepackage{arxiv}

\usepackage[utf8]{inputenc} 
\usepackage[T1]{fontenc}    
\usepackage{hyperref}       
\usepackage{url}            
\usepackage{booktabs}       
\usepackage{amsfonts}       
\usepackage{nicefrac}       
\usepackage{microtype}      
\usepackage{lipsum}
\usepackage{pgfplotstable}
\usepackage{graphbox,graphicx}
\usepackage{ulem}
\usepackage{tabularx}
\usepackage{rotating}
\usepackage{caption}

\title{Social distancing with the Optimal Steps Model}

\author{
  Christina Maria Mayr \thanks{See also \texttt{www.vadere.org}} \\
  Munich University of Applied Sciences\\
  Department of Computer Science and Mathematics\\
  Lothstrasse 64 \\
  80335 Munich \\
  \texttt{christina\_maria.mayr@hm.edu} \\
   \And
 Gerta Koester \\
  Munich University of Applied Sciences\\
  Department of Computer Science and Mathematics\\
  Lothstrasse 64 \\
  80335 Munich \\
  \texttt{gerta.koester@hm.edu} \\
}

\begin{document}


\maketitle

\begin{abstract}
With the Covid-19 pandemic, an urgent need has arisen to simulate social distancing. The Optimal Steps Model (OSM) is a pedestrian locomotion model that operationalizes an individual's need for personal space. We present new parameter values for personal space in the OSM to simulate social distancing in the pedestrian dynamics simulator \textit{Vadere}. 
Our approach is pragmatic. We consider two use cases: in the first, we demand that a set social distance must never be violated. {In the second the social distance can be violated temporarily for less than $10s$.} For each use case we conduct simulation studies in a typical bottleneck scenario and measure contact times, that is, violations of the social distance rule.
We conduct regression analysis to assess how the parameter choice depends on the desired social distance and the corridor width. 
{We find that evacuation time increases linearly with the width of the repulsion potential, which is an analogy to physics modeling the strength of the need for personal space. The evacuation time decreases linearly with larger corridor width. The influence of the corridor width on the evacuation time is smaller than the influence of the range of the repulsion, that is, the need for personal space.} If the repulsion is too strong, we observe clogging effects.  
Our regression formulas
enable \textit{Vadere} users to conduct their own studies without understanding the intricacies of the OSM implementation and without extensive parameter adjustment.
\end{abstract}



\section{Introduction}
In 2020, distance rules were imposed in many countries to slow down the spread of the coronavirus.
Obeying such rules leads to a change in crowd behavior: pedestrians keep more distance~\cite{caspar-2020-cdyn} and, to achieve this, might slow down their walking speed
{ 
\cite{echeverria-huarte-2021-cdyn}. 
But what happens when people do not have enough space to keep a distance from each other? For example at a train station or a narrow bottleneck in an evacuation situation.  Bottlenecks have been investigated in numerous studies under normal conditions.  
In three laboratory experiments~\cite{seyfried-2009-cdyn,tian-2012-cdyn, ren-2019-cdyn}, the authors find that the flow depends linearly on the corridor width. 
Ren et al.~\cite{ren-2019-cdyn} assess the evacuation time of elderly people through a bottleneck from a laboratory experiment. They find that the total evacuation time depends piecewise linearly on the exit width.
Zhang et al.~\cite{zhang-2019b-cdyn} conduct an egress experiment with mice. They find that the total evacuation time decreases with the increase of the exit width in a certain range. They also find that the mean flow rate shows a nonlinear dependency on the exit width.

Locomotion behavior has changed due to social distancing measures. This poses a challenge to existing movement models that are calibrated to normal behavior. According to~\cite{ronchi-2020b-cdyn}, several crowd simulation frameworks have released new features to incorporate social distancing in their simulation models\footnote{Unfortunately, the authors do not refer to any simulation framework directly. We could spot  e.g. Pathfinder (https://support.thunderheadeng.com/docs/pathfinder/2020-3/user-manual/) and Pedestrian Dynamics (
https://www.incontrolsim.com/simulate-social-distancing-with-pedestrian-dynamics/). Both accessed on 14th September 2021))}.

Typically models that describe the movement of crowds are categorized as macroscopic, mesoscopic, and microscopic~\cite{cristiani-2014-cdyn}. 
Microscopic models are advantageous when investigating distance behavior, since they model each person as an individual. The distances between agents can thus be measured directly. This is not possible with meso- or macroscopic models, as only density and flow values are available\footnote{To investigate social distancing indirectly, the density could be related to a distance value. E.g. results from laboratory experiments show that  a social distance of 1m is guaranteed if the density is below 0.16ped/m2 ~\cite{echeverria-huarte-2021-cdyn}. However, such results are not generally available.}.

Microscopic models can be divided into non-differential and differential models~\cite{cristiani-2014-cdyn}. The group of force based models belongs to the differential models. They use the concept of force equilibrium from mechanics~\cite{cristiani-2014-cdyn}. Agents and obstacles are modeled as rigid bodies between which there are attractive and repulsive forces. A disadvantage of this type of model is unrealistic oscillations~\cite{cristiani-2014-cdyn, koster-2013-cdyn}. If the repulsive forces are too strong, agents keep advancing and regressing. Probably the most widely used force model is the social force model~\cite{helbing-1995-cdyn}, for which numerous extensions~\cite{johansson-2007-cdyn, yu-2005-cdyn, lakoba-2005-cdyn, parisi-2009-cdyn, ratsamee-2012-cdyn, yang-2021-cdyn} exist. 

As far as the authors know, the social force model has not yet been calibrated for social distancing. However, in the context of infection studies in which social force models were extended by SIR models, there are first attempts to achieve social distancing by adjusting the social force model parameters. Harweg et al.~\cite{harweg-2021-life} vary the model parameter in a social force model that controls how far repulsive forces affect other agents and measure the exposure time. Sajjadi et al.~\cite{sajjadi-2021-life} control the parameter social distancing intensity and measure the exposure risk that is the fraction of agents exposed to the infection. Neither Harweg et al.~\cite{harweg-2021-life} nor Sajjadi et al.~\cite{sajjadi-2021-life} assess how the distancing depends on the respective parameter value. Only Sajjadi et al.~\cite{sajjadi-2021-life} demonstrate that changing the respective parameter leads qualitatively to the expected distancing behavior. Whether the desired distance is achieved by the choice of the parameter value remains unclear. This is a problem because the parameters have no physical meaning. In addition, the change of a single parameter can result in several effects~\cite{kretz-2018-cdyn, dias-2018-cdyn}. 

Cellular automata constitute another type of microscopic model. Cellular automata are non-differential and are mainly used in large-scale simulations~\cite{lubas-2016-cdyn}. The principle of operation is simple: the area is divided into cells. Each agent moves from cell to cell, with different probabilities assigned to the cells using, for example, a multi logit model~\cite{lubas-2016-cdyn}. In most cases, each agent occupies exactly one cell. The advantage is simplicity and speed, the drawback is 
a coarse spatial discretization. Dealing with the discretization is a challenge for calibration~\cite{kirchner-2004-cdyn, lubas-2016-cdyn, dias-2018-cdyn}, even without fine granular effects like social distancing. Lovreglio et al.~\cite{lovreglio-2015-cdyn} suggest to adjust the parameters of the multinomial logit model that is used to determine the probability values. One parameter weights the influence of the static floor field in the multinomial model, the second parameter that of the dynamic floor field. The parameter values of the floor fields are combined in one~\cite{dias-2018-cdyn} or more~\cite{lovreglio-2015-cdyn} parameter vectors. In this case, the parameters are abstract, that is, not directly measurable. Despite the large number of studies on the qualitative validation of cellular automata~\cite{kretz-2009-cdyn, dias-2018-cdyn}, there are only a few studies in which a cellular automaton was calibrated against empirical data~\cite{dias-2018-cdyn, dias-2019-cdyn, taherifar-2019-cdyn}. As far as the authors know, no cellular automaton has yet been calibrated for social distancing.

Another type of non-differential microscopic models are optimal steps models~\cite{seitz-2012-cdyn,sivers-2014-cdyn,sivers-2015-cdyn, sivers-2016-cdyn, zeng-2018b-cdyn}.  They combine the step-wise movement of cellular automata with the continuity of space of social force models~\cite{seitz-2012-cdyn}. Agents set their steps exactly where the value of a utility function attains its maximum at a certain point in time~\cite{seitz-2012-cdyn}. The utility function balances the conflicting goals of reaching a destination and keeping a distance from other agents and obstacles.
It consists of a static part, the floor field, which encodes the (negative) geodesic distance to a target while skirting obstacles,
and a dynamic part, which takes into account other agents that must be avoided. This is done through an analogy from physics similar to the one in force-based models: agents carry a repulsion potential that translates into a reduced utility for others.  
First approaches\footnote{https://www.accu-rate.de/wp-content/uploads/DISTANSIM-technical-report.pdf} already show that optimal steps models can represent social distancing by a suitable choice of parameters.

So far, none of the studies have investigated the relationship between parameters and distance behavior. This is important because the distance rules differ from country to country. The World Health Organization recommends a distance of 1m~\cite{who-2021-life}, while the U.S. Centers for Disease Control and Prevention recommend six feet (1.82m)~\cite{center-us-2021-life}. In Australia  and Germany, a distance of 1.5m is recommended~\cite{australia-2021-life, rki-2021d-life}, while, in the UK, the distance should be at least 2m~\cite{uk-2021-life}. Existing locomotion models should therefore be able to represent different distance behaviors under pandemic conditions. We believe that this could be realized with optimal steps models. Our research question is therefore: 
}

\begin{itemize}
\item How can one adjust parameters in optimal steps models to achieve physical distancing that reflect specific desired \lq social\rq{} distances?
\end{itemize}


In this study, we try to find parameter values that allow us to simulate social distancing. For that purpose, we will use the implementation of the optimal steps model in the free open-source pedestrian dynamics simulator \textit{Vadere}\footnote{see \url{https://www.vadere.org/} and~\cite{kleinmeier-2019-cdyn}}.
In the following, we will denote with capital letters the OSM or OSM as implemented in \textit{Vadere}, while small letters refer to the model class.

We look at a classic bottleneck scenario. We argue that this scenario is particularly interesting because many geometries in the built environment can be interpreted as a sequence of bottlenecks. 
We also consider different desired social distances, which we define through the Euclidean distance between the centers of the agents. Agents in \textit{Vadere} are represented by circles. We call  \textit{contact} any violation of the desired social distance, even for the shortest time. 

We investigate two use cases: \\
\begin{itemize}
\item Use case 1: Social distance as a lower bound. In this use case, the social distance value must never be violated. As a consequence, if we find such a parameter combination, there are no contacts at all. In other words, we interpret the social distance as a lower bound of the actual distances kept by the agents. 
\item Use case 2: Social distance as target distance. In this use case, we accept a temporal violation $(\leq10s)$ of the
distance requirement, that is, contacts. 
\end{itemize}

For both use cases, agents need to keep larger distances than before. In the OSM, personal distance is modeled  by introducing \lq repulsion\rq{} between agents, or less physically interpreted, the utility of
a position decreases when an agent approaches another agent. The repulsion (or utility drop) is captured by a so-called potential function. Hence, we need to  find appropriate parameters of the potential function to strengthen the repulsion between agents.

The paper is organized as follows. In Section 2, we introduce the OSM, the scenario, and define criteria that represent the two use cases. In Section 3, we propose a new methodology to assess the relationship between desired social distance, OSM parameters, and the corridor width of the bottleneck. In addition, we assess how the evacuation time is affected by these parameters. We discuss under which conditions the effect of clogging occurs and we propose strategies to tackle this issue. Section 4 provides a conclusion.

\begin{figure}
\centering
\includegraphics[width=1.0\textwidth]{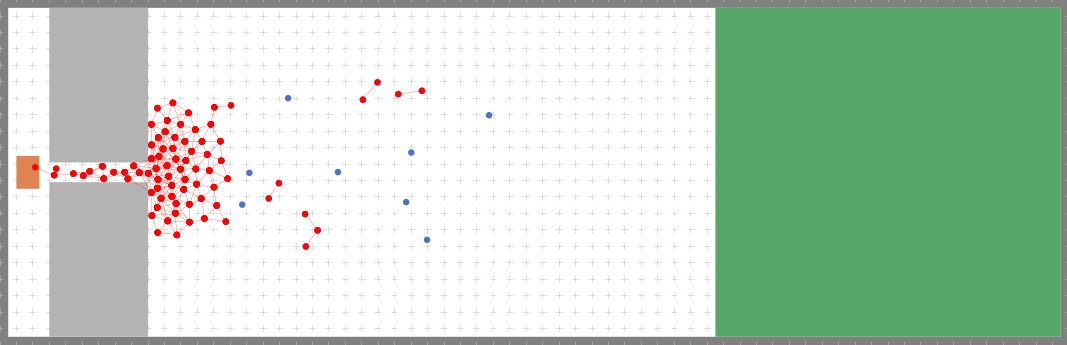}
\caption{Normal crowd behavior in front of a bottleneck. The agents (circles) try to reach the target on the left-hand side. In this example, we consider agents being in contact if they violate a social distance of $1.5m$. Red circles represent agents currently in contact. The contact between two agents is represented by a red connection line.  Blue circles represent agents out of contact. \textit{Vadere's} default parameters for the OSM are chosen to reflect such normal behavior. For a simulation with social distancing new parameters must be found.}
\label{fig:normalCrowd}
\end{figure}

\section{Materials and methods}
\subsection{The OSM}
The basic idea of the OSM is that virtual pedestrians (agents) are attracted by targets and repulsed by obstacles and other virtual pedestrians. Less physically spoken, agents, when moving, maximize the utility of their position. This utility depends (negatively) on the geodesic distance to the target and (negatively) on the proximity to other agents. Agents move by \lq stepping\rq{} on the position within a circle around their current spot that optimizes this utility. The circle radius represents each agent's personal maximum stride length,
which in turn is linearly correlated to the agent's free-flow speed~\cite{seitz-2012-cdyn}, that is, an assumed desired speed when the path is free. Thus, agents step towards targets while skirting obstacles and avoiding collisions. In the remainder of the text, we use the physical interpretation, because it is used in the names of parameters in  \textit{Vadere's} implementation of the OSM. 

{
The potential $P_l(x)$ for pedestrian $q$ at time $t$ for an arbitrary point $x \in \mathbb{R}^2$ in the  plane is defined as~\cite{seitz-2012-cdyn}
\begin{equation}
P_l(x) = P_t(x,s) + \sum_{j=1}^m P_{o,j}(x) + \sum_{i=1,i\neq l}^n P_{p,i}(x) 
\end{equation} 
where $P_t(x,s)$ is the attraction potential of the target, $s$ is the vector of the current agent positions, $P_{o,j}(x)$ is the repulsive potential induced by obstacle $j$, $m$ is the number of obstacles, $P_{p,i}$ is the repulsive potential of other agents, $n$ is the number of agents. The implementation of the attraction potential $P_t(x)$ is the so-called floor field. If the attraction potential does not depends on the current agents' positions, means $P_t(x,s)=P_t(x)$, the floor field is static. Otherwise, it is dynamic.
We look at static floor fields. In this case, the value of the potential function at position $x$ equals the negative geodesic distance between $x$ and the target.

The obstacle potential $P_{o,j}(x)$ is defined as 
\begin{eqnarray}
P_{o,j}(x) = \left\{
\begin{array}{ll}
\delta_1(x) + \delta_2(x) & d_{o,j}(x) < r_l \\
\delta_1(x)  & r_l \leq d_{o,j}(x) < w_o \\
0 & \, \textrm{otherwise.} \\
\end{array}
\right.
\end{eqnarray}

where $d_{o,j}(x)$ is the distance to the closest point of the obstacle from position $x$, $w_o$ defined the width of the obstacle repulsion, $r_l$ is the torso radius of agent $l$. The functions $\delta_i$ are defined as
\begin{eqnarray}
\delta_1(x) = 6 \exp{\left[ 2 \left( \left(  \frac{d_{o,j}(x)} {w_o} \right)^{2} - 1\right)^{-1} \right]} \\
\delta_2(x) = 10^5 \exp{\left[ \left( \left(  \frac{d_{o,j}(x)} {r_l} \right)^{2} - 1\right)^{-1} \right]} \\
\end{eqnarray}

The repulsion between two agents is achieved by the distance-dependent potential function $P_{p,i}$. In Vadere, it is implemented as
\begin{eqnarray}
P_{p,i}(x) = \left\{
\begin{array}{ll}
\phi_1(d_{l,i}(x)) + \phi_2(d_{l,i}(x)) + \phi_3(d_{l,i}(x)) & d_{l,i}(x) <r_i + r_l \\
\phi_1(d_{l,i}(x)) + \phi_2(d_{l,i}(x)) & \, r_i + r_l \leq d_{l,i}(x) < w_{int}  +r_i+r_l \\
\phi_1(d_{l,i}(x))  & \, w_{int} + r_i + r_l \leq  d_{l,i}(x) < w+r_i+r_l\\
0 & \, \textrm{otherwise.} \\
\end{array}
\right.
\label{eq:potentialAgentNew}
\end{eqnarray}
where $d_{l,i}(x)$ is the Euclidean distance between agent $l$  and agent $i$. $r_i, r_l$ are the respective torso radi of agent $i,l$. $w_{int}$ and  $w$ are the intimate and the personal space width according to Hall. The functions $\phi_i$ are defined as
\begin{eqnarray}
\phi_1({d_{i,j}}) = \mu \exp{\left[ 4 \left( \left(  \frac{d_{i,j}} {w +r_i +r_l} \right)^{2} - 1\right)^{-1} \right]} \\
\phi_2({d_{i,j}}) = \frac{ \mu } {a_p} \exp{\left[ 4 \left( \left(  \frac{d_{i,j}} {w_{int} +r_i +r_l} \right)^{2} - 1\right)^{-1} \right]} \\
\phi_3({d_{i,j}}) = 10^3 \exp{\left[ \left( \left(  \frac{d_{i,j}} {r_i +r_l} \right)^{2} - 1\right)^{-1} \right]}
\end{eqnarray}
}

\begin{figure}[hbt!]
\centering
\includegraphics[width=11cm]{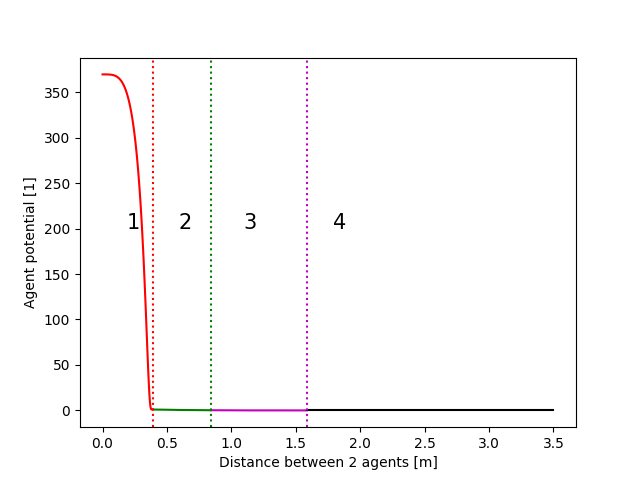} 
\includegraphics[width=11cm]{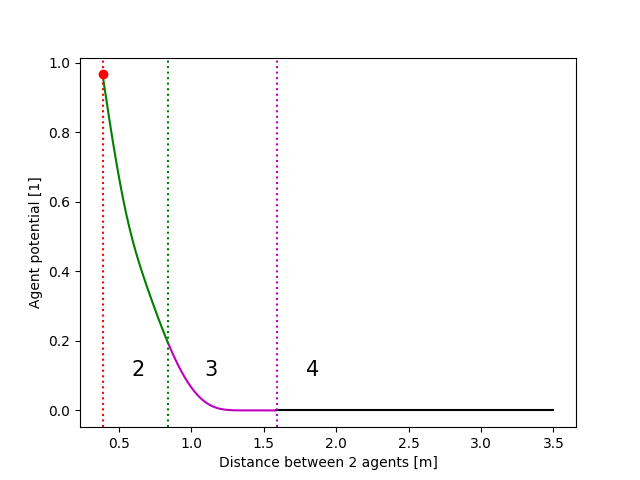} 

\caption{{Agent potential function, or repulsive potential, according to Eq.~\ref{eq:potentialAgentNew} for default parameter values \textit{potential height} $h=50$  and \textit{personal space width} $w=1.2$. Note: the scale of the ordinates differ in the sub-figures.} The personal space is 
around the torso of an agent. If the distance between two agents is smaller than the sum of the torso radii, the agents collide (1). To prevent this, the potential is set to a high value in the torso area. The personal space (2) begins at $x=0.84m$ and ends at $x=1.59m$. {The upper bound $x=1.59m$ is the sum of the personal space width $w=1.2$ and the torso radii of the agents. (3) represents the personal space. (4) is a region out of interest.}}
\label{fig:standard}
\end{figure}

See Figure~ \ref{fig:standard}. The potential function is based on Hall's theory of interpersonal distances which describes four distance zones around a person~\cite{sivers-2016-cdyn, hall-1966-life}. 
Accordingly, the potential function is defined piece-wise on rings around each agent: a circular core for collision avoidance, a first ring that represents the intimate space, and a second ring that represents personal space.
Agents outside the personal zone do not affect other agents' path choice. This is mathematically modeled by setting the  the potential function to zero.  
 
The value of the potential function in the personal space ring is very low. Thus, this area will be kept free only if  agents have ample space to avoid each other~\cite{sivers-2016-cdyn}. As soon as the space becomes more 
constricted agents will get closer. This is typical for normal human behavior. In the intimate space ring , 
the potential function value increases significantly. Again see Figure~\ref{fig:standard}.
In crowds, this area is only kept free if the density is low ~\cite{sivers-2016-cdyn}. 
Finally, to prevent agents from overlapping, the potential is set to a very high value (when compared to the values in the personal and intimate spaces) in the collision area. The exact definition of the potential function, and default parameters, implemented in \textit{Vadere} can be found in~\cite{kleinmeier-2019-cdyn}.

The shape of the potential function is controlled through several parameters that could be adapted to achieve social distancing in the OSM. We try to keep the adaption of the OSM as simple as possible by only changing two parameters: the \textit{potential height} $h$ (in \textit{Vadere}: pedPotentialHeight) and the \textit{personal space width} $w$ (in \textit{Vadere}: pedPotentialPersonalSpaceWidth). 
Why do we choose these two parameters? The parameter \textit{potential height} $h$  controls the strength of repulsion. If $h$ is increased, we expect agents to increase their distance to others. 
The \textit{parameter personal space width} $w$ controls how far the repulsion reaches:
the larger $w$ the bigger the influence area of an agent.

Note that the \textit{personal space width} $w$ is related to but not equal to
the desired social distance $d$ the user wants to model. When there is ample space it might suffice to set $w=d$ to keep agents at least the desired social distance apart.
Since the true distance between agents is an emergent value, we expect the distance to be larger. In a bottleneck scenario, like Figure~\ref{fig:normalCrowd} on the other hand, the true distance will be much smaller than $w$.  

In crowd simulations, we are especially interested in dense crowds which typically occur in front of bottlenecks, such as doors or narrow passages. As a consequence, 
if we want to observe a virtual crowd that obeys social distancing, 
we need to choose a value for the personal space width $w$ that is bigger than the desired social distance.  For this purpose, we want to find suitable values for the parameters \textit{personal space width} and \textit{potential height}.

\begin{figure}[hbt!]
\centering
\includegraphics[width=0.2\textwidth, align=bottom]{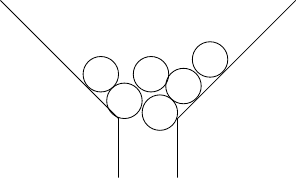}
\hspace{1.5cm}
\includegraphics[width=0.2\textwidth, align=bottom]{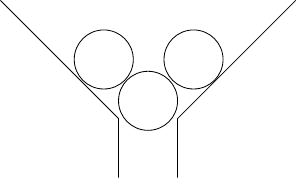}
\hspace{1.5cm}
\includegraphics[width=0.2\textwidth, align=bottom]{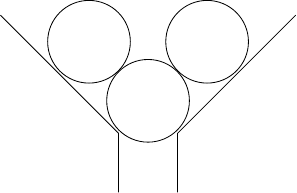}

\caption{Funnel with balls. The ball size represents the repulsion between agents. If the ball size is small (small repulsion), the balls just fall through the narrowing. If the ball size is too large, balls get stuck. This is why we must not increase the repulsion to an infinite value to achieve social distancing. }
\label{fig:funnel}
\end{figure}

\subsection{Scenario}

There is currently no empirical data available to calibrate the OSM parameters for social distancing. Nonetheless, we need to find a suitable scenario and define plausible criteria which help us to evaluate how well social distancing is captured by the model.

{When selecting the scenario, the following aspect should be considered}. In use case 1, the social distance is a lower bound for permissible distances. In particular, all distances larger than the social distances are accepted. One might be tempted to achieve such behavior by increasing the repulsion between agents up to \lq infinity\rq, that is, to set \textit{the potential height} $h$  and the \textit{personal space width} $w$ to extremely high values. 

{Such a procedure can cause the effect of clogging.} If the potential is too high, agents get stuck, as illustrated in Figure \ref{fig:funnel} with a funnel analogy. The funnel represents a bottleneck that agents need to pass to reach their target. The ball size represents the repulsion between agents. If the ball size is small (small repulsion), the balls just fall through the narrowing. If the ball size is too large, balls get stuck. 
This effect can be easily observed in test simulations. {We conclude} that we cannot increase the repulsion at will. 

We use the bottleneck scenario depicted in Figure \ref{fig:normalCrowd} with this consideration in mind. The bottleneck scenario covers both aspects, density and repulsion. We can directly observe what happens if the repulsion is too high. 

The scenario is set up as follows. The topography of the scenario is $12m$ wide and $65m$ long. 
The length of the corridor is $5.0m$. We consider ten discrete corridor widths $c \in \{1.2m, 1.4m, 1.6m, ... , 3.0m\}$. 100 agents are generated in the source on the right-hand side (green box) and try to reach the target (orange box) on the left-hand side. With these settings, the agents are close to each other in front of the bottleneck. The source is placed sufficiently far away from the corridor to exclude any effects of the spawning process on the observation area. We start to count contacts at simulation time $t=20s$ to exclude any contacts produced by the spawning process. The distance measure between two agents is the Euclidean distance between their center points.

\subsection{Choice of criteria for acceptable parameters}

We do not have any empirical data to calibrate our parameters for social distancing. Hence, we need other criteria to decide which parameter values fit our two use cases best.  
%
{We look at the original motivation for social distancing:}
to decrease the number of contacts and with it the probability of infection. 
{With this in mind, we select} a time measure $t_m$ for counting contacts {as proposed in~\cite{ronchi-2020b-cdyn}}. 

The time measure $t_m$ is defined as
\begin{equation}
t_m = \frac{1}{n}\sum_{i=0}\sum_{j=0}{t_{i,j}}
\end{equation}
In our investigation,  the number of agents is $n=100$. 
$t_{i,j}$ is the time agent $i$ has contact to another agent $j$. The time $t_{i,j}$ is zero if the distance $x_{i,j}$ between two agents $i$ and $j$ is above the desired social distance: 
\begin{equation}
t_{i,j}=\left\{
    \begin{array}{ll}
      0, & \mbox{if $x_{i,j} > d$}.\\
      t_{s,e} - t_{s,s}, & \mbox{otherwise}.
    \end{array}
  \right.
\end{equation}
where $t_{s,s}$ is the point of time when the social distance is violated for the first time and $t_{s,e}$ is the point of time when the contact ends.

Adding up the durations of social distance violations may seem less intuitive than direct distance measures at a first glance. However, evaluating the relative impact of distances can become very complex, since we do not know how much a certain distance matters.

For use case 1, the contact time has to be zero. This means that all agents always keep a distance larger than the desired social distance $d$. Thus, the acceptable parameter combinations for use case 1 have to fulfill the condition:
\begin{equation}
t_m = 0 \quad \mbox{Condition for use case 1}
\label{eq:cond1}
\end{equation}

We expect that this condition is always fulfilled when the \textit{personal space width} and the \textit{potential height} are set to high  values which leads to high repulsion. Since a high repulsion can lead to clogging, we only want to increase the repulsion as much as necessary. We expect that there are multiple parameter combinations that fulfill condition 1. Hence, the solution to our problem is not unique but forms an indifference curve. 

For use case 2, we accept parameter combinations which fulfill: 
\begin{equation}
t_m \leq 10s \quad \mbox{Condition for use case 2}
\label{eq:cond2}
\end{equation}



\section{Results and discussion}

\subsection{Methodology for finding suitable parameter values}
There are three parameters in the simulation, the personal space width $w$, the corridor width $c$, and the personal space height $h$. These influence the distance behavior and thus the evacuation time. The parameters personal space width and space height can be freely chosen, while the corridor width is conditioned by the topography. The aim is to find out which parameter values have to be chosen to achieve use cases 1 and 2 for a given social distance and how this is related to the evacuation time. 

We find that both use cases can be covered by adjusting the personal space width while fixing the personal space height at $h=850$, see Appendix \ref{sec:fixing}.
In our studies, we vary the parameters personal space width and corridor width, and keep the personal space height at $h=850$. We use ten corridor widths $c \in \{1.2,1.4,1.6, .., 3.0\}$. For the personal space width, we choose a finer discretization $w \in \{1.200,1.225,1.250, ..5.0\}$.

This resolution is necessary to determine, with sufficient accuracy, those parameter values for which 
our acceptance thresholds $(t_m=0s, t_m=10s)$ are passed. 

In summary, we run $10$x$150$ simulations and collect the contact times for the social distances $1.25m,1.50m,1.75m,2.0m$. {For each use case and for each social distance, we find the minimum values of the personal space width for which the respective use case and the respective distance behavior is fulfilled: $w_{min,uc=1}, w_{min,uc=2}$.  We also collect the evacuation time for each simulation run.  

\subsection{Relationship between personal space width and social distancing}
To fulfill the condition for use case 1 and 2, the personal space width $w$ must be selected depending on the social distance $d$, that one wishes to achieve, and on the corridor width $c$, see Fig.~\ref{fig:pngregressionSpace1},\ref{fig:pngregressionSpace2}. 
We observe that the personal space width $w$ depends linearly on the desired social distance. This is in line with our expectations. For each use case, we set up a linear model (least squares) using the data points from Fig.~~\ref{fig:pngregressionSpace1},\ref{fig:pngregressionSpace2}.
The independent variables are the values of the respective personal space widths $w_{min,uc}$ values and the values of the corridor widths. In case of outliers (samples with a studentized residual $|r_{s,i}|\geq 3$), we remove the corresponding samples and repeat the regression procedure.
The coefficients of the linear model are depicted in Tab.~\ref{tab:useCase1Regression}.
In both use cases, the personal space width $w$ depends on the corridor width $c$ and the desired social distance $d$. Both, $d$ and $c$ are statistically significant, see Tab.~\ref{tab:useCase1Regression},\ref{tab:useCase2Regression}. The regression formula for use case 1 is:
\begin{eqnarray}
\label{eq:rule1w}
w = 2.555d-0.1170c-1.4521 \\ 
\label{eq:rule1h}
h = 850 \textrm{ (const.)}  \\ 
d \in [1.25,2.00],\quad c \in [1.2,3.0] \nonumber
\end{eqnarray}
The regression formula for use case 2 is:
\begin{eqnarray}
\label{eq:rule2w}
w = 1.6444d-0.0658c-0.6161 \\ 
\label{eq:rule2h}
h = 850 \textrm{ (const.)}  \\ 
d \in [1.25,2.00],\quad c \in [1.2,3.0] \nonumber
\end{eqnarray}

It is problematic that the distance behavior depends on topography parameters such as the corridor width, as these are usually not freely chosen. 
There may even be several corridors with different widths in a simulation. 
This problem is alleviated by the comparatively small effect of the corridor width. The two upper confidence intervals are even close to zero, see Tab.~\ref{tab:useCase1Regression},\ref{tab:useCase2Regression}.
Also, 
the coefficients of the social distances are at least twenty times higher than the coefficients of the corridor width so that the result changes only slowly with changing corridor widths. This effect can also be observed in Figures \ref{fig:pngregressionSpace1}, \ref{fig:pngregressionSpace2}. Note that this argument is only valid because the parameter ranges are similar.  For practical applications, we argue, that the value of the corridor width does not need to be matched exactly. In the case of multiple corridor widths, we recommend using the minimum corridor width. In this case, the agents would maintain slightly higher distances in broader corridors. 
\begin{minipage}{\textwidth}
\centering
\includegraphics[width=11cm]{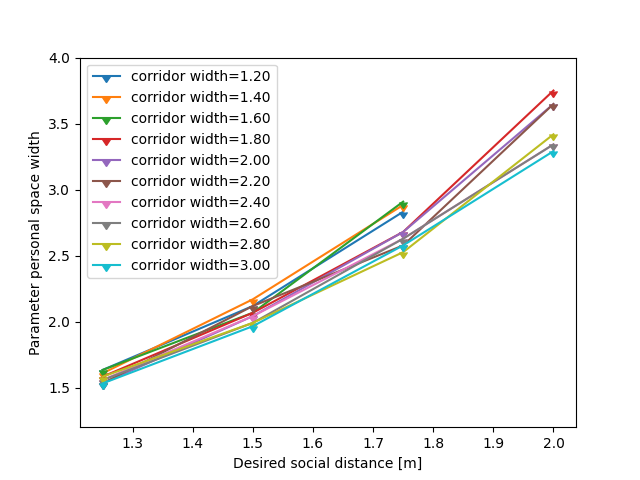} 
\captionof{figure}{Use case 1. The personal space width depends on the desired social distance and the corridor width. In case of missing data points, no personal space width could be found that fulfils the criteria of use case 1. The personal space width is set to $h=850$.}
\label{fig:pngregressionSpace1}
\begin{tabular}{@{}lllllll@{}}
\toprule
              & coef    & std err & t-stat       & p-value & {[}0.025 & 0.975{]} \\  \midrule
1        & -1.4521 & 0.134   & -10.859 & 0.000              & -1.817   & -1.087   \\
Social distance $d$     & 2.5550  & 0.075   & 34.084  & 0.000              & 2.350    & 2.760    \\
Corridor width $c$ & -0.1170 & 0.036   & -3.245  & 0.003              & -0.215   & -0.019  \\ \bottomrule
\end{tabular}
\captionof{table}{Use case 1. Regression results for the personal space width. The adjusted coefficient of determination $R^2=0.972$, $pValue=0.000$, sample size = 37.}
\label{tab:useCase1Regression}
\end{minipage}

\begin{minipage}{\textwidth}
\centering
\includegraphics[width=11cm]{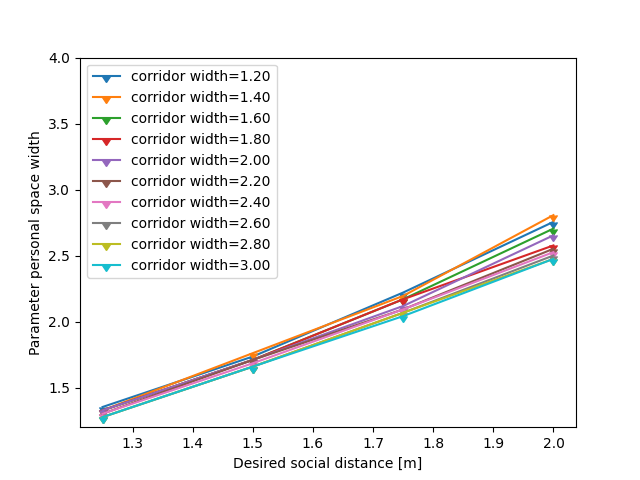} 
\captionof{figure}{Use case 2. The personal space width depends linearly on the desired social distance. The effect of the corridor width slightly increases with increasing social distance. However, the desired social distance seems to be the most influential parameter.The personal space width is set to $h=850$.}
\label{fig:pngregressionSpace2}
\centering
\begin{tabular}{@{}lllllll@{}}
\toprule
              & coef    & std err & t-stat      & p-value & {[}0.025 & 0.975{]} \\  \midrule
1         & -0.6161 & 0.022   & -27.634 & 0.000              & -0.662   & -0.571   \\
Social distance $d$     & 1.6444  & 0.013   & 128.675 & 0.000              & 1.618    & 1.670    \\
Corridor width $c$ & -0.0658 & 0.006   & -11.043 & 0.000              & -0.078   & -0.054  \\ \bottomrule
\end{tabular}
\captionof{table}{Use case 2. Regression results for the personal space width. The adjusted coefficient of determination $R^2=0.998$, $pValue=0.000$, sample size = 36.}
\label{tab:useCase2Regression}
\end{minipage}

\subsection{Evacuation time}
The evacuation time depends on the corridor width $c$ and the personal space width $w$, see Fig.~\ref{fig:evacTimes}. With increasing personal space width, the evacuation time increases. With increasing corridor width the evacuation time decreases. In some cases, temporary clogging occurs, which means, that agents get stuck for a certain amount of time. These cases correspond to the peeks in Fig.~\ref{fig:evacTimes}. The cases of stationary clogging (evacuation time $t_{evac}\geq1500s$), are not depicted in Fig~\ref{fig:evacTimes}.

We conduct a Least Squares regression to assess how the evacuation time depends on the parameter personal space width and social distance. The independent variables are the personal space width and the corridor width. The dependent variable is the evacuation time. Samples where the evacuation time is infinity (stationary clogging) and outliers (samples with a studentized residual $|r_{s,i}|\geq 3$), are removed from the regression procedure. The resulting regression formula is
\begin{eqnarray}
\label{eq:ruleEvac1}
t_{evac} = 92.5272w-74.5817c+194.8345 \\ 
w \in [1.2,5.0], c \in [1.2,3.0] \nonumber
\end{eqnarray}

\begin{minipage}{\textwidth}
\centering
\includegraphics[width=12cm]{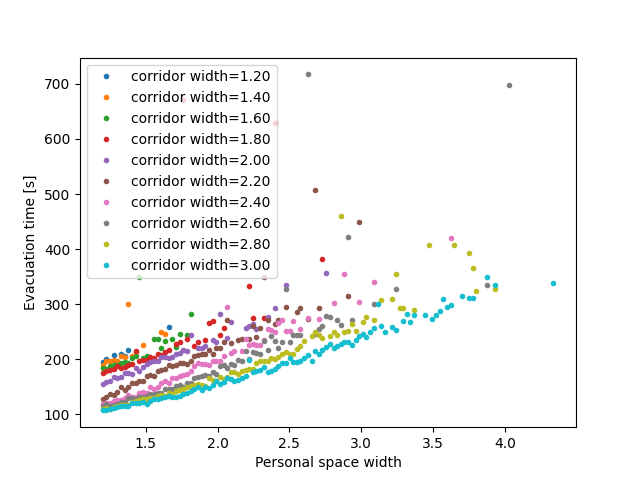}
\captionof{figure}{Evacuation time. The evacuation time increases over the personal space width. High evacuation times (outliers) result from the effect of temporary clogging. Outliers are removed in the regression procedure. Not that in case of missing data points, the evacuation time was infinite due to clogging effects.} 
\label{fig:evacTimes}
\begin{tabular}{@{}lllllll@{}}
\toprule
Parameter                             & coef     & std err & t-stat       & p-value & {[}0.025 & 0.975{]} \\ \midrule
1                          & 194.8345 & 2.466   & 79.013  & 0.000              & 189.988  & 199.681  \\
Corridor width $c$                & -74.5817 & 1.119   & -66.664 & 0.000              & -76.780  & -72.383  \\
Personal space width $W$ & 92.5272  & 0.852   & 108.602 & 0.000              & 90.853   & 94.202   \\ \bottomrule
\end{tabular}
\captionof{table}{Regression results. Evacuation time. The adjusted coefficient of determination $R^2=0.964$, $p-Value=0.000$, sample size = 448. The personal space width is set to $h=850$.}
\label{tab:evacTimeRegressionResults}
\end{minipage}
The results of the regression can be found in Tab.~\ref{tab:evacTimeRegressionResults}. The regression coefficient of the parameter personal space width supports our observation that the evacuation increases with increasing personal space width. The corresponding regression coefficient is $92.5272$. The coefficient of the corridor width $(-74.5817)$ shows that  with increasing corridor width, the evacuation time linearly decreases. 

These results are plausible and extend the findings of~\cite{seyfried-2009-cdyn,tian-2012-cdyn, ren-2019-cdyn}. If we have to keep larger distances, fewer people can enter the corridor at the same time. They have to enter one after the other, which costs more time and increases the evacuation time. The larger the corridor width, the more people can enter the corridor at the same time. 

With the help of Eq.~\ref{eq:rule1w},~\ref{eq:rule2w},~\ref{eq:ruleEvac1}, it is possible to qualitatively estimate how much longer an evacuation time will take if the distance recommendation is increased. For this purpose, the personal space width is first calculated as a function of the personal distance and the corridor width according to Eqs.~~\ref{eq:rule1w},~\ref{eq:rule2w} and then used in Eq.~~\ref{eq:ruleEvac1}.

\subsection{How clogging is related to social distancing}
\label{sec:clog}
We want to assess under which condition clogging occurs. In this study, we assume clogging when the evacuation time is larger than the simulation time: $t_{evac}\geq1500s$. We expect that clogging occurs when the personal space width $w$ is larger than the corridor width $c$: $w > c$. Clogging is more likely when the repulsion potential is more dominant than the target potential. Suppose two agents are standing in front of a bottleneck. One agent stands at the left corner, the other at the right corner. In the case that the repulsion potential is smaller than the corridor width, the agents only see the target potential. In the other case, the agents see the repulsion potential of the other agent, which is superposed with the target potential. This is more likely with increasing range and height of the  repulsion potential. 

For each simulation, we collect the minimum values of the personal space for which clogging occurs (evacuation time $t_{evac} > 1500s$, $1500s$ = max. simulation time). We want to assess whether the condition $w\geq c$ holds for clogging. For that purpose, we compute the Pearson correlation coefficient and plot minimum values of the personal space over the corridor with. We find that $c$ and $w$ are highly correlated (Pearson correlation coefficient = $0.977$). Moreover, the borderline values are close to the unit line that represents the exact condition $w = c$, see Fig.~\ref{fig:clogging}. We find that the condition for clogging $w\geq c$ broadly holds.

\begin{figure}
\includegraphics[width=12cm]{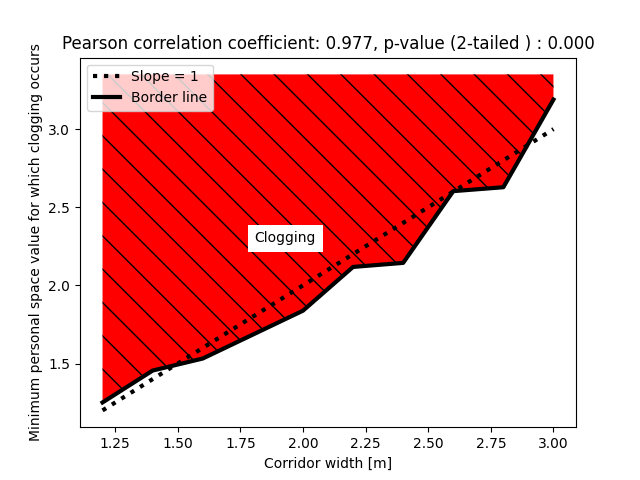} 
\caption{Occurrence of clogging.}
\label{fig:clogging}
\end{figure}

To avoid clogging in the simulation, the personal space width values should be smaller than the width of the corridor. Otherwise, one has to expect clogging. For practical applications, where the corridor width is set we suggest the following strategies.
The first strategy is to extend the locomotion behavior by a suitable  behavioral model.
We think, that real pedestrians resolve deadlocks by temporarily adapting their behavior. For example, they might briefly violate social distances to \lq squeeze by\rq{}. }
Another strategy would be to model a time-dependent agent potential function, where the 
repulsion decreases when agents experience a certain time of clogging. This
would entail social distance violations and, in our eyes, model pedestrians' impatience with deadlocks. Such a behavior is already implemented in the \textit{accu:rate's} simulator \textit{crowd:it}\footnote{\url{https://www.accu-rate.de/en/}}. After a certain time, agents switch to their default behavior, that is, without keeping any distance rules. This usually resolves the clogging. We plan to implement similar behaviors in the \textit{Vadere} simulation framework.

For the time being, we recommend keeping this problem in mind when using our parameter values in \textit{Vadere}. We strongly recommend using a dynamic floor field and allow agents to switch places in counterflows. Both measures decrease the probability of clogging. 
See Tab.~\ref{tab:additionalSettings} for the necessary \textit{Vadere} settings.

\begin{figure}
\centering
 \includegraphics[width=0.45\textwidth, trim=0 0 400 0,clip]{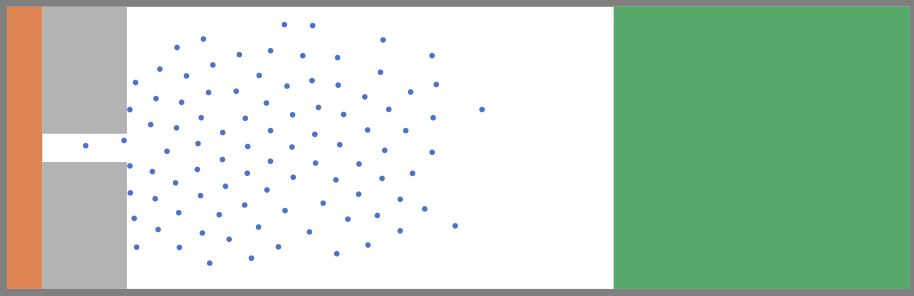} 
\hspace{1cm}
 \includegraphics[width=0.45\textwidth, trim=0 0 400 0,clip]{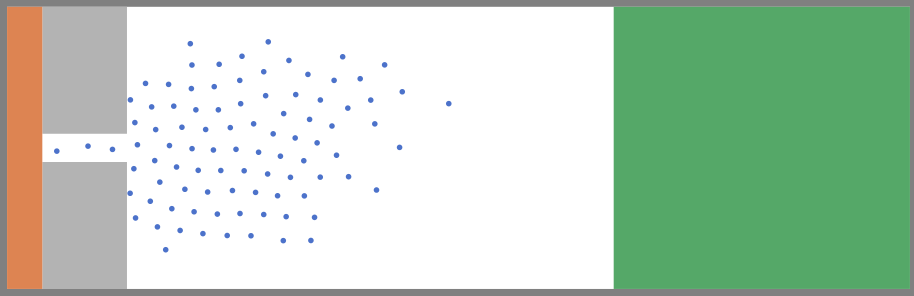} 

\caption{Crowd behavior for use case 1 (left) and 2 (right) at simulation time $t_s=40s$ (corridor width $c=2.0m$).  For use case 1, the parameter values are set to $w(d=1.5m,c=2.0m)=2.1464, h=850$ (see Eq.~\ref{eq:rule1w} ). No contacts are allowed $(t_m=0s)$. In use case 2 (right), a temporary violation $(t_m=10s)$ of the desired social distance is allowed. In this case, the parameters are set according to Eq.~\ref{eq:rule2w}: $w(d=1.5m,c=2.0m)=1.7189, h=850$. One can observe that agents need to keep more distance in use case 1 than in use case 2, because the contact time requirement is harder to fulfill.}
\label{fig:averagecase}
\end{figure}

\begin{table}[hbt!]
\centering
\begin{tabularx}{\textwidth}{@{}llll@{}}
\toprule
Use case    & \textit{Vadere} parameter   & Default & Value for social distancing       \\ \midrule
Use case 1   & pedPotentialPersonalSpaceWidth  $w$ & 1.2    &   see Equation (\ref{eq:rule1w})              \\
            & pedPotentialHeight $h$   & 50.0    & 850  \\
Use case 2  & pedPotentialPersonalSpaceWidth $w$  & 1.2    & see Equation (\ref{eq:rule2w})              \\
             & pedPotentialHeight $h$   & 50.0    & 850 \\ \bottomrule
\end{tabularx}
\caption{Overview of parameters to achieve social distancing with the OSM.}
\label{tab:Parametertable}
\end{table}

\begin{sidewaystable}
\begin{tabular}{@{}lllll@{}}
\toprule
Reduce probability of clogging        & \multicolumn{2}{l}{Vadere parameter}  & Default & Value for social distancing        \\ \midrule
in unidirectional flows                & Floor field & timeCostAttributes/type & UNIT    & NAVIGATION                  \\
in counterflows & Psychology  & usePsychologyLayer      & false   & true                        \\
                                       &             & searchRadius            & 1.0     & 1.5 desired social distance \\
                                       
&& cognition & SimpleCognitionModel  &CooperativeCognitionModel \\                                       \bottomrule
\end{tabular}
\captionof{table}{Overview of additional settings in \textit{Vadere} to reduce the probability of clogging.}
\label{tab:additionalSettings}
\end{sidewaystable}

\subsection{Limitations and open research questions}
Currently, the distancing behavior is defined by a time measure. However, the precision of the equations could be increased by reformulating the problem as an optimization problem where each deviation from the desired social distance is punished within a utility function. For this purpose, several questions need to be answered: In which area around an agent should a distance deviation be punished? How to choose the utility function? The solution to this problem is part of future research. 

The {regression formulas} are based on a bottleneck scenario where high densities occur. We expect that our parameter suggestions work for scenarios with bottlenecks. We can not guarantee that they work for any topography. 

When testing the new parameter values, we observed that the probability of clogging increases when increasing the social distance. This is caused by the higher repulsion. The clogging problem can be mitigated but not eliminated by using a dynamic floor field
and allowing position switches in counterflow. We list the \textit{Vadere} parameter settings to achieve this. However, we believe that the problem is better addressed by implementing suitable agent behaviors. We plan to do this in the \textit{Vadere} simulation framework.
{
In the present study, only one simulation run per sample, i.e. one fixed seed, was used due to the limited computing capacity. It would be interesting to see what results are obtained with several simulation runs per sample. }

\section{Conclusions}

{

We investigated how to calibrate parameters of the OSM, an optimal steps model implemented in the free and open-source \textit{Vadere} framework so that social distancing can be simulated.
We considered two use cases for social distancing: In the first, contacts had to be completely avoided. The distance between agents could not fall below the required social distance value. In the second use case, a temporal violation of the social distance of $10s$ on average per agent was allowed. For both use cases, we derived regression formulas to determine the OSM parameters for any desired social distance $d \in [1.25m, 2.00m]$. 

Our procedure was based on two steps.
In the first step, we varied the OSM parameters personal space height and personal space width, that describe an agent's need for personal space, for a narrow $(1.2m)$ and broad $(3.0m)$ bottleneck. We analyzed the contact duration for each of 2500 samples. We found that the conditions for both use cases could be fulfilled with personal space height of 850 and kept it fixed at that value in the following.

Then, we investigated, for both use cases, how the personal space width has to be chosen depending on the desired social distance and the corridor width. We found that the personal space width depends linearly on both, the desired social distance and on the corridor width. However, the latter has a significantly smaller influence. We also saw that evacuation time increases linearly with the personal space width and decreases linearly with the corridor width. 

We observed that agents get stuck in some simulations. We found that clogging always occurs in our bottleneck scenario when the personal space extends beyond the corridor width.  

Several questions remain to be answered. The first is the problem of clogging for which new sub-models need to be introduced that resolve unrealistic clogging in the simulation where real persons would find a cooperative solution to proceed. 
The second is the definition of the use cases that are currently based solely on time measures, that is, on the duration of the violation of a fixed distance.
Measures that take into account how much the distance was violated, have the potential to be much more
 precise. However, it is unclear, how to weigh these distances. 
 Another  open question is to what extent the formulas can be transferred to other scenarios than the considered bottleneck. 
}

{With our findings, optimal steps models, notably the OSM as in \textit{Vadere}, can be calibrated to the changed locomotion behavior under social distancing measures. Crowd flows under pandemic conditions can be simulated more realistically. Furthermore, we better understand how the evacuation time depends on the social distancing behavior. }

\section*{Acknowledgments}

We thank Dr. Angelika Kneidl and Alex Platt (B.\ Sc.) from \textit{accu:rate} who provided insight and expertise that greatly assisted the research. In particular, we thank for the stimulating discussions and the testing of the parameters in real-life applications.  

\bibliographystyle{unsrt}  

\bibliography{Literatur}

\appendix

\section{Fixing the parameter personal space width}
\label{sec:fixing}
The question is whether the personal space height and width have to be adjusted at the same time, or whether adjusting one parameter is sufficient.  To test this, we vary the two parameters at the upper and lower bound of the corridor width: $c_{lb}=1.2m$ and  $c_{ub}=3.0m$. We assume a linear behavior in between.

We consider the social distances $d$: $1.25, 1.5, 1.75, 2.0$. For each, we determine the indifference curves which fulfill the conditions as defined in Equations \ref{eq:cond1}, \ref{eq:cond2}.  

\begin{figure}[hbt!]
\centering
\includegraphics[width=7cm]{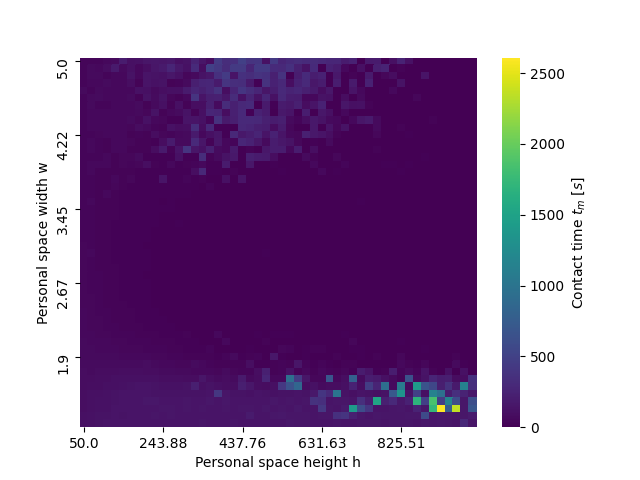} 
\includegraphics[width=7cm]{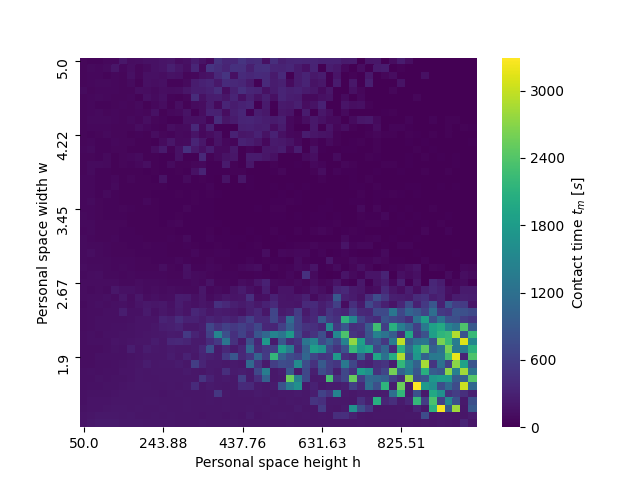} 
\caption{Average contact time (third dimension, see colorbar) over \textit{potential height} $h$ and \textit{personal space width} $w$ for a social distance $d=1.5m$ (left) and $d=2.0m$ (right) at a fixed corridor width $c=1.2m$. In $h \in [250,825]$, the contact time increases again over $w$. In this area, pairing occurs, because the agent potentials are low but reach very far. As a result, very many agent potentials are superposed at the same time. The next minimum in the step circle radius can then happen to be next to the neighbouring agent, causing random pairing. We want to avoid such behavior and require that $h\geq850$.
}
\label{fig:surfacewc12}
\end{figure}

\begin{figure}[hbt!]
\centering
\includegraphics[width=7cm]{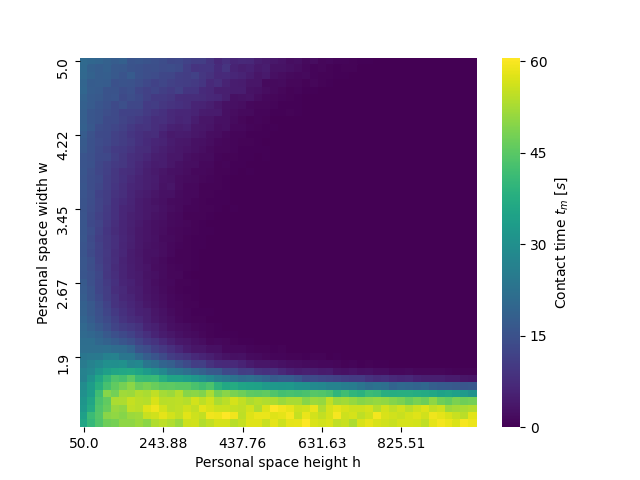} 
\includegraphics[width=7cm]{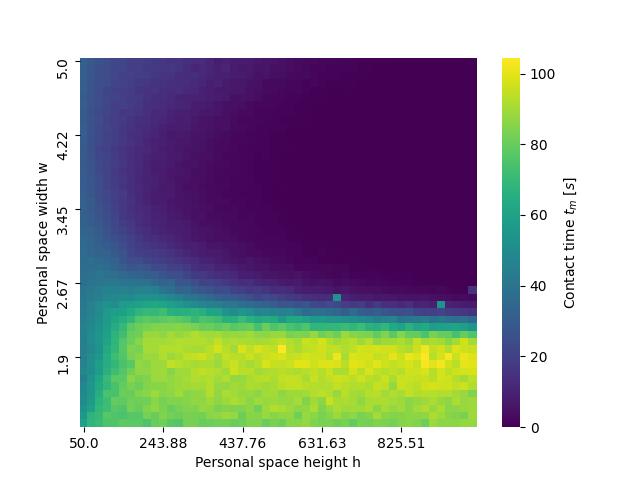} 
\caption{Average contact time (third dimension, see colorbar) over \textit{potential height} $h$ and \textit{personal space width} $w$ for a social distance $d=1.5m$ (left) and $d=2.0m$ (right) at a fixed corridor width $c=3.0m$. For $h \in [250,400]$, pairing occurs, because the agent potentials are low but reach very far. As a result, very many agent potentials are superposed at the same time. The next minimum in the step circle radius can then happen to be next to the neighbouring agent, causing random pairing. We want to avoid such behavior and require that $h\geq850$.
}
\label{fig:surfacewc30}
\end{figure}

For that purpose, we use a simple grid sampling. Each parameter is discretized with 50 values which are equally spaced between lower and upper bound. See Table \ref{tab:parameter}. In total, we get 2500 parameter combinations. For each of these parameter combinations, we run a simulation. 
The average contact times for each use case are depicted in Fig.~\ref{fig:surfacewc12},\ref{fig:surfacewc30}. We find that the contact average time decreases when increasing the potential height $h$ or the personal space width $w$ except for $h \in [250,825]$ (Fig.~\ref{fig:surfacewc12}) and $h \in [250,400]$ (Fig.~\ref{fig:surfacewc30}). In these cases, the contact time increases over $w$ due to random pairing that is caused by low  agent potentials that reach very far. As a result, many agent potentials are superposed at the same time. The next minimum in the step circle radius can then happen to be next to the neighboring agent, causing random pairing. We want to avoid such behavior and require that $h\geq850$.

In the next step, we analyze the resulting average contact time $t_m$ for each of the four social distances  $d$: $1.25, 1.5, 1.75, 2.0$.

\begin{figure}
\includegraphics[width=7cm]{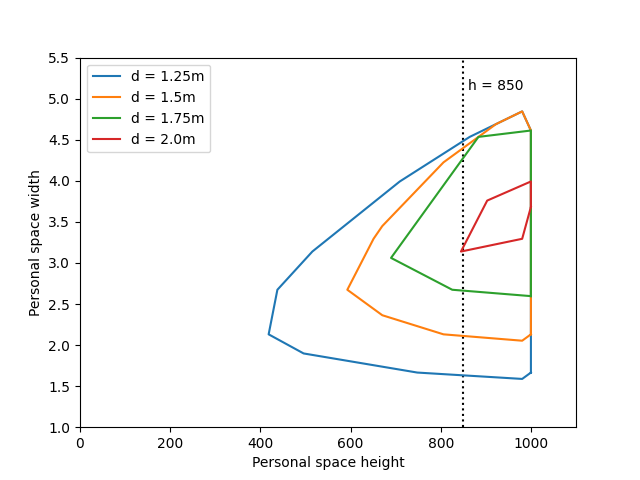} 
\includegraphics[width=7cm]{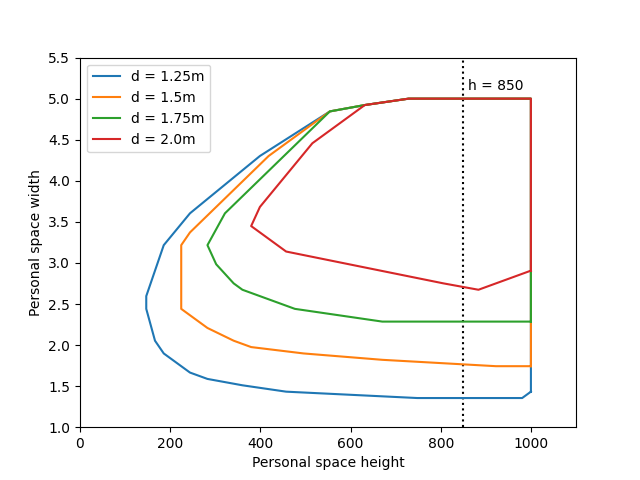} 
\caption{Indifference curves for use case 1 (left) and use case 2 (right) at corridor width $c=1.2m$ (lower bound). In use case 1, contacts are not allowed. Each indifference curve (left) fulfills the condition $t_m=0$. The indifference curves (right) correspond to an average contact time $t_m=10s$. Both use cases can be fulfilled for a fixed personal space height $(h=850)$, because the $h=850$-line intersects each indifference curve.}
\label{fig:paretofronts1}
\end{figure}

\begin{figure}
\includegraphics[width=7cm]{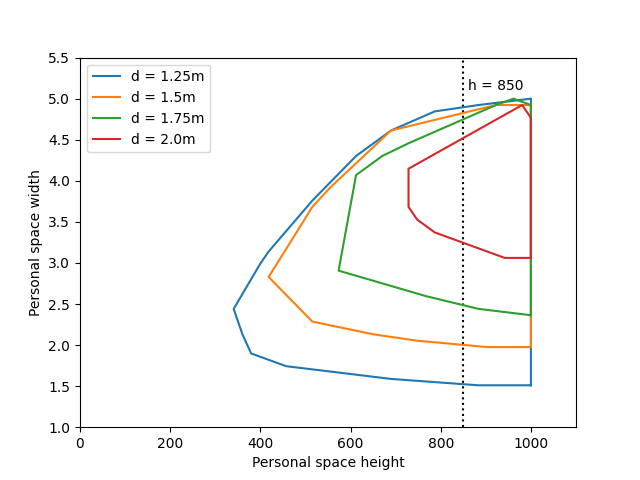} 
\includegraphics[width=7cm]{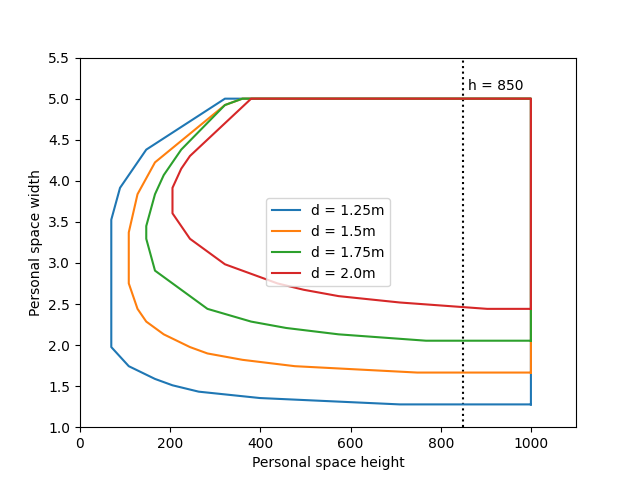} 
/\caption{Indifference curves for use case 1 (left) and use case 2 (right) at corridor width $c=3.0m$ (upper bound). In use case 1, contacts are not allowed. Each indifference curve (left) fulfills the condition $t_m=0$. The indifference curves (right) correspond to an average contact time $t_m=10s$. Both use cases can be fulfilled for a fixed personal space height $(h=850)$, because the $h=850$-line intersects each indifference curve.}
\label{fig:paretofronts2}
\end{figure}

We start with use case 1. For each social distance $d_i$, we mark all parameter combinations which fulfill $t_m=0$. Then we build the convex hull around these points which serves as an approximation for the indifference curve. 
We proceed analogously with use case 2 ($t_m=10s$). The indifference curves for the corridor widths $c_i=\{1.2m,3.0m\}$ are depicted in Fig.~\ref{fig:paretofronts1},\ref{fig:paretofronts2}.
 
We see that both use cases at a fixed personal space height $h = 850$, see Fig.~\ref{fig:paretofronts1},\ref{fig:paretofronts2}. It is therefore not necessary to adjust both space width and height at the same time. For this reason, we fix the personal space height to $h=850$, in the following.

\begin{table}[hbt!]
\centering
\begin{tabular}{@{}lcccc@{}}
\toprule
Parameter                 & Default & Lower bound & Upper bound & Number of discrete values \\ \midrule
Personal space width $w$ & 1.2                                                                                & 1.2         & 5.0         & 50            \\
Potential height $h$     & 50.0                                                                               & 50.0        & 1000        & 50             \\ \bottomrule
\end{tabular}
\caption{Analyzed parameter values. Each parameter is discretized by 100 values which are equally spaced between lower and upper bound. In summary, this leads to 10000 parameter combinations. }
\label{tab:parameter}
\end{table}

\end{document}